# KEY FEATURES OF ADMINISTRATIVE RESPONSIBILITY

**Vladimir Zhavoronkov,**
*Associate Professor, Russian University of Transport,*
*Moscow, Russia*

**Valeri Lipunov,**
*Associate Professor, Russian University of Transport,*
*Moscow, Russia*

**Mattia Masolletti,**
*Lecturer, NUST University,*
*Rome, Italy*

**Abstract**

The article examines both the legal responsibility itself and its types, and in various aspects. The authors apply legal analysis, as well as the principles of consistency and integrity. The contradictions of administrative responsibility, as well as legal gaps in its interpretation, are highlighted.

**Key words:** administrative responsibility.

**JEL codes:** K-1; K-4; K-10.

## 1. Introduction

Despite the seriousness of recent research in the field of administrative responsibility, some aspects still require further study. The term 'administrative responsibility' can be found in 26 articles of the Administrative Code of the Russian Federation, as well as in the title of its second chapter (the current version). It should be noted that, despite the importance of the institution of administrative responsibility, there is no normative definition that allows to reveal its content and essence.

## 2. Main part

The importance of defining such aspects of administrative responsibility is that they allow us to assess the regularity of administrative responsibility, create conditions for its improvement and the possibility of comparing it with other types of legal responsibility.

Thus, 'retrospective (negative) legal liability (including administrative liability) is considered as a form (mechanism) of the state's response to an offense; as the obligation of the offender to undergo measures of state coercion provided for by the sanction for committing an offense and the implementation of this obligation; as the application of punishment provided for by law to persons who have committed an offense; as a negative state assessment and condemnation of the offense and the person who committed it, accompanied by the application of punishment' [1]. All these judgments, on the one hand, reflect administrative responsibility. The content of this legal relationship is 'the right of the state (represented by authorized bodies and officials) to apply measures of state (administrative) coercion to a person who has committed an administrative offense, and the obligation of the latter to undergo deprivation and the right to restrictions associated with these measures [2].

Administrative responsibility is the ability of the body endowed with the appropriate competence (employee) to apply the due sanction to the subject who has committed an administrative offense. Administrative responsibility, being a kind of legal responsibility, has signs peculiar to legal responsibility, but it also has special ones. So, for instance: administrative penalties can be contained in both laws and legal regulations. Therefore, administrative responsibility has its own legal basis. The basis for the application of the sanction by the body is the commission of a misdemeanor. Both individual and collective entities are held accountable. It follows from the above that responsibility and misconduct are related phenomena that are connected by a causal relationship, and also that an administrative offense is an independent institution of administrative and tort law [3].

Material and procedural norms are combined in the institute of administrative and tort law, which raises questions, since their subject and method of legal regulation do not coincide. For example, material norms are aimed at mandatory regulation of relations, the subjects of which are only the state (represented by its bodies and officials) and the person who committed an administrative offense, and procedural norms regulate the relations between these participants and victims, witnesses, experts and other participants in the proceedings, while dispositive methods are also used (for instance, the possibility of some participants in the proceedings to refuse to testify, file petitions, present evidence, appeal the decision in the case).

Consequently, the substantive and procedural institutions of administrative responsibility can be considered as interdependent, but at the same time relatively independent components of the institute of administrative and tort law. "The material and legal part consists of the norms of the General Part of the Administrative Code of the Russian Federation, it consists of principles such as: equality before the law of persons who have committed administrative offenses (Article 1.4 of the Administrative Code), the legality of the application of administrative punishment (Article 1.6 of the Administrative Code), as well as the norms defining the procedure for the operation of legislation on administrative offenses in time and space (Articles 1.7 and 1.8 of the Administrative Code), establishing the features of administrative responsibility of certain subjects (Articles 2.4 - 2.6, 2.10), defining the goals and types of punishments, the rules for their appointment, the statute of limitations for bringing to administrative responsibility and the period of stay of a person subjected to administrative punishment in a state of 'administrative punishment', the possibility of releasing a person who has committed an administrative offense from administrative responsibility (Chapters 3 and 4 of the Administrative Code of the Russian Federation)' [3]. In addition, the sanctions of the norms of the Special Part of the Code specify the types and sizes (terms) of administrative penalties that are established by the General Part. The procedural institute of administrative responsibility consists of norms aimed at regulating the proceedings in cases of administrative offenses, as well as the execution of decisions in cases of administrative offenses (sections IV, V of the Administrative Code of the Russian Federation).Since the procedural institute is a legal one, it consists of sub-institutes, such as the institute of evidence- in Chapter 26 of the Administrative Code, the institute of initiating an administrative offense case- in Chapter 28, as well as the consideration of the case- in Chapters 29, 30, and 31. In addition, it includes the norms of the General Part of the Code (Articles 1.5 and 1.6) and in Chapter 24 (Articles 24.2 and 24.3) and others. For proper regulation, it is necessary to use both the substantive and procedural institutions of administrative responsibility in their unity. Each institute has its own gaps that require due consideration.

Let us consider the following problems of the administrative responsibility:

- *The uncertainty of the place of administrative responsibility in the system of responsibility.*

Retired judge of the Constitutional Court of the Russian Federation Vitruk noted in one of his works that 'an administrative offense and a criminal offense are homogeneous in their social nature, and therefore both an administrative offense and a criminal offense require punishments that differ only in the severity of negative consequences for the corresponding offenders' [4].

The consistent practice of the European Court of Human Rights, in which the concept of the 'criminal prosecution' is considered 'autonomously', also indicates the unified nature of these illegal actions. The States parties to the European Convention on Human Rights are deprived of the opportunity to remove certain types of offenses from the scope of the Convention by simply changing their names. While generally agreeing with this approach, it should be noted that at present the boundaries between administrative offenses and criminal offenses are being erased. Falling into a 'prohibitive rage', the legislator constantly expands the scope of administrative coercion. Gradually, a situation is developing in which it becomes simply impossible to draw a clear line between administrative and criminal responsibility from the point of view of the severity of the alleged punishment. And the blurring of the boundaries of responsibility of various types generates injustice of the most unpredictable forms and properties. For instance, when smaller adverse consequences correspond to greater procedural guarantees.

- *The differentiation of administrative responsibility.*

Back in 1961, the Italian jurist Bruno Leoni came to the conclusion that the main obstacle to the implementation of the rule of law is excessive legislation [5].

Recently, the federal legislator has been trying to differentiate administrative responsibility more and more, without obvious necessity, multiplying the legislation and using rather controversial methods for this – for instance, fixing the place of its commission as a sign of an offense-the territory of a particular subject of the Russian Federation. And if for 'compositions of action' there may still be some dispute regarding the discrimination of this feature, then for 'compositions of inaction', for instance, part 3 of Article 18.8 of the Administrative Code of the Russian Federation, the presence of discrimination is obvious, since the place of commission of such an offense according to established judicial practice will be considered the place of residence of the person being held accountable (subparagraph 'h' of paragraph 3 of the Resolution of the Plenum of the Supreme Court of the Russian Federation 'On some issues arising from courts when applying the Administrative Code of the Russian Federation' No. 5, adopted on March 24, 2005).

- *Instability of the legislation on administrative responsibility.*

It is important to say that only in 2015, more than 50 federal laws had been adopted, which amended the Administrative Code of the Russian Federation. Without assessing the quality of these changes, only relying on their number is enough to state the existence of a problem, namely, legal instability. As Shuvalov rightly notes, 'the abundance of changes made to laws, the rapid replacement of adopted norms make it difficult to inform those to whom these laws are addressed', and, ultimately, generates legal nihilism. Giving preference to the 'amendment approach', the legislator turns even once harmonious legal prescriptions into a kind of patchwork quilt-a chaotic mixture of old and new rules, often contradicting each other. We should fully agree with the judge of the Constitutional Court of the Russian Federation, Mr. Yaroslavtsev, that the words from the Holy Scripture sound more and more modern: 'And no one puts patches of unbleached cloth to old clothes, because the newly sewn will tear away from the old, and the hole will be even worse' [1].

## 3. Conclusion

An administrative offense is an illegal action that contradicts the norms of law specified in the laws, it can be expressed as actions or inaction. At the same time, the subject must understand what he is doing and realize the inevitability of the onset of negative consequences –a strong-willed character. The purpose of the administrative rules is aimed at maintaining state discipline and public order, including on the roads, when it comes to transport security.

**References**

[1] Khamanev N. Administrative Law of the Russian Federation M., 2015. p. 326.

[2] Galagan I. A. Administrative responsibility in the USSR (state and material-legal research). - Voronezh, 1970. p. 51

[3] Shergin A. P. Conceptual foundations of administrative-tort law // Scientific portal of the Ministry of Internal Affairs of Russia. 2008. No. 1. p. 18.

[4] Vitruk N. V. General theory of legal responsibility. 2nd edition, corrected and supplemented. M., 2009. p. 129.

[5] Leonie B. Freedom and the law; trans. from the English. V. Koshkin. M., 2008. p. 307.